\documentclass[a4paper,11pt]{article}
\usepackage{pos}
\usepackage{subcaption}

\usepackage[draft,inline,nomargin]{fixme} \fxsetup{theme=color}
\FXRegisterAuthor{cc}{acc}{\color{OliveGreen}CC}
\FXRegisterAuthor{hlv}{hlv}{\color{red}HLV}

\usepackage{siunitx}
\usepackage{physics}
\usepackage{parskip}       
\hypersetup{
  colorlinks=true,
  linkcolor=blue,
  citecolor=blue,
  urlcolor=blue,
}
\usepackage{lineno}

\title{Search for seasonal variations of the horizontal muon rate with the HAWC observatory}
 \ShortTitle{Seasonal variations of the horizontal muon rate}

\author[]{Cindy Castellón}
\author*{Hermes León Vargas}
\onbehalf{on behalf of the HAWC collaboration}

\affiliation{Instituto de Física, Universidad Nacional Autónoma de México\\
  Mexico City, México}

\emailAdd{cindym@estudiantes.fisica.unam.mx}
\emailAdd{hleonvar@fisica.unam.mx}

\abstract{The High-Altitude Water Cherenkov (HAWC) observatory was designed to study gamma-ray sources in the energy range between a few hundred GeV up to few hundred TeV. It is composed of 300 Water Cherenkov Detectors (WCDs) that cover a surface of approximately $\SI{22000}{\meter\squared}$, at $\SI{4100}{\meter}$ a.s.l. In this study, we use the HAWC WCDs as a very large horizontal particle tracker, searching for horizontal muon rate variations by season using 1.5 years of HAWC data. We look for a possible correlation between the effective temperature and the horizontal muon rate. In order to do this, we developed a method to calculate the effective temperature for the horizontal propagation of muons. This is the first time that a search for seasonal variations in the high-altitude horizontal muon rate is reported.}

\ConferenceLogo{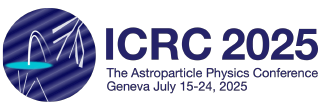}

\FullConference{39th International Cosmic Ray Conference (ICRC2025)\\
 15–24 July 2025\\
Geneva, Switzerland\\}

\begin{document}
\maketitle
\section{Introduction}
The High-Altitude Water Cherenkov (HAWC) observatory, located at an altitude of \SI{4100}{\meter} in Puebla, Mexico, consists of 300 water Cherenkov detectors (WCDs), each equiped with four photomultiplier tubes (PMTs), distributed over an instrumented area of \SI{22000}{\meter\squared}. While HAWC was designed to detect gamma rays in the energy range from hundreds of GeV to hundreds of TeV, it can also detect horizontal muons produced high in the atmosphere, these reach the detector after traveling long paths, being affected by the physical conditions of the upper atmosphere. Previous work by the HAWC collaboration explored the possibility of detecting Earth-skimming neutrinos using nearly horizontal muon-like tracks~\cite{hawc-neutrinos}. This analysis builds upon the data selection and reconstruction techniques developed for that work but it focuses on the horizontal muon rates as a function of atmospheric conditions.

The dependence of the vertical muon flux on atmospheric temperature is well studied, mainly by underground experiments~\cite{minosfar,verpoest2024}. However, in the horizontal case, this dependence is not yet known. Because of the long atmospheric paths, horizontal muon production and propagation may exhibit different sensitivities to temperature variations compared to vertical fluxes. In this work, we combine experimental data from HAWC, atmospheric modeling based on real data, and numerical simulations performed with the Matrix Cascade Equations (MCEq) software~\cite{fedynitch2015} to investigate this possible correlation. We present the methods used to calculate the effective atmospheric temperature relevant for horizontal muon production, the simulations performed under different conditions and hadronic interaction models, and a direct comparison with the experimental data collected over more than one year of detector operation.

\section{Horizontal Muons with HAWC}

This analysis builds upon the data selection and reconstruction framework developed for the HAWC Earth-skimming neutrino search, reported in a previous ICRC edition~\cite{hawc-neutrinos}. The main difference lies in the use of an expanded dataset that includes all available high-quality data spanning more than one year, along with stricter requirements on detector stability to enable reliable long-term rate studies. The dataset consists of triggered events reconstructed with extended information, including detailed timing and charge distributions from the PMTs in each WCD.

\begin{figure}[h]
\begin{subfigure}[t]{0.49\textwidth}
\centering
\includegraphics[width=\textwidth]{ 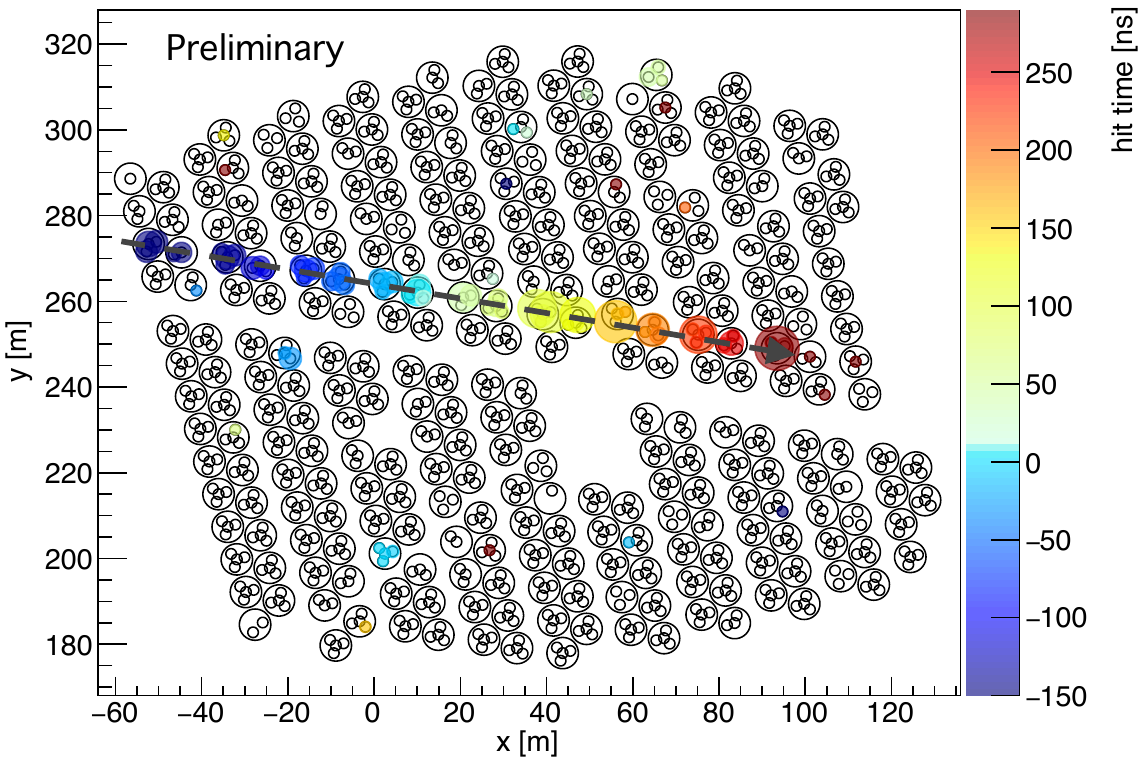}
\end{subfigure}
\hfill
\begin{subfigure}[t]{0.49\textwidth}
\centering
\includegraphics[width=\textwidth]{ 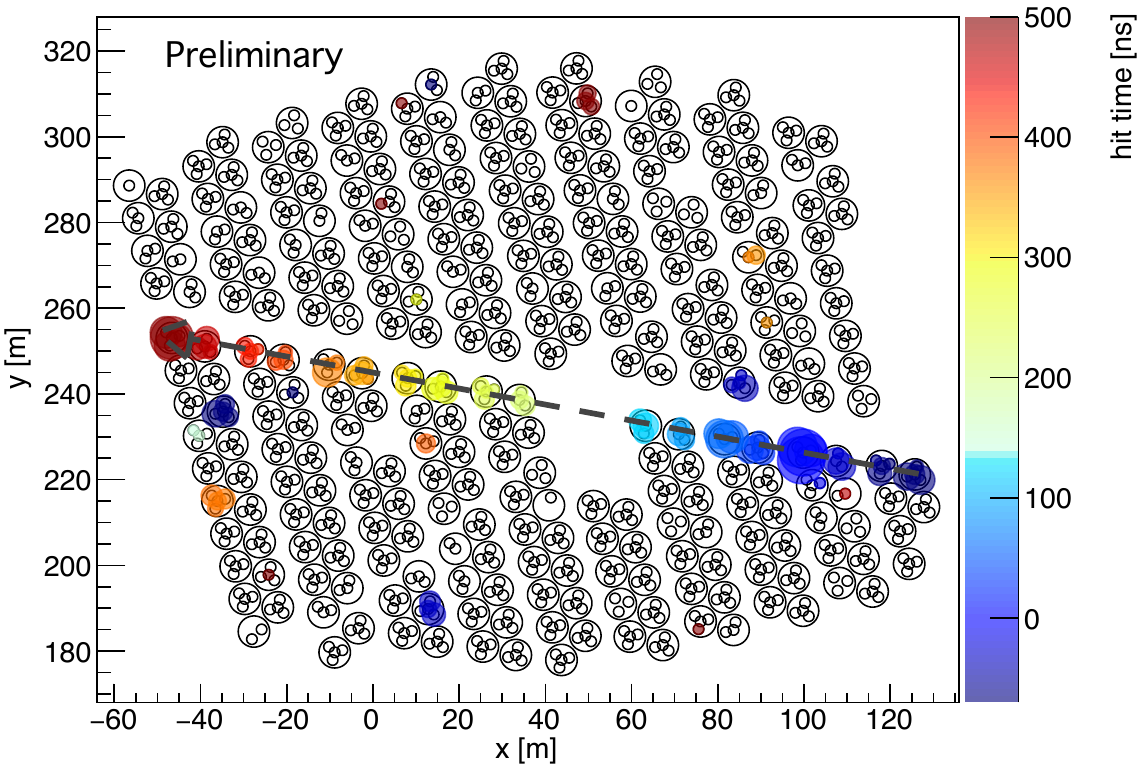}
\end{subfigure}
\caption{Event displays of horizontal muon candidates in HAWC. Color is showing arrival time, consistent with the pattern expected for a near-horizontal trajectory. We consider two opposing directions; from southeast and from northwest.}
\label{fig:event-display}
\end{figure}

\begin{figure}[h]
\centering
\includegraphics[width=0.6\textwidth]{ 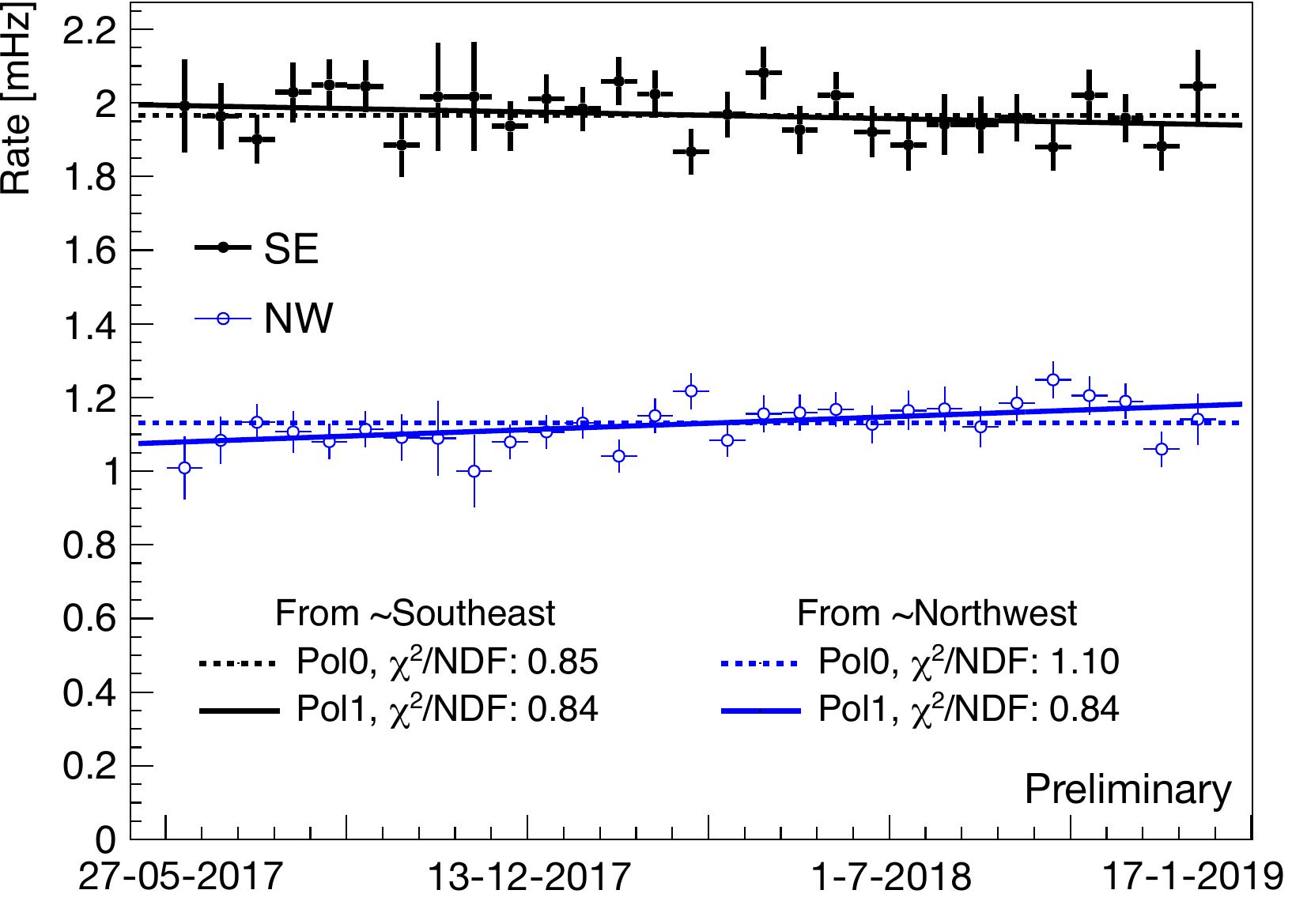}
\caption{Time variation of the horizontal muon rate over 1.5 years of HAWC data.}
\label{fig:rates-hawc}
\end{figure}

Since HAWC is surrounded by mountains, this study focuses on a direction where the horizon is unobstructed, allowing muons to reach the detector nearly horizontally. Figure~\ref{fig:event-display} shows an example of a selected horizontal muon event. Although the array's geometry is not fully symmetric, the chosen direction aligns with a long row of detectors, enhancing detection efficiency. The corresponding detection rate of horizontal muon candidates over time is shown in Figure~\ref{fig:rates-hawc}. For both opposing directions (approximately southeast and northwest), the rate remains approximately constant over the whole time window (from may 2017 to january 2019), with no clear evidence of seasonal modulation. We also include fits to a constant function and a first-degree polynomial; the constant fit yields a $\chi^2/\mathrm{NDF}$ between 0.85 and 1.10, supporting the interpretation of a stable rate.

\section{Effective temperature calculation}
The atmospheric data used for our analysis is from the European Centre for Medium-Range Forecasts (ECMWF)~\cite{europeancentre}. The ECMWF data combines measurements from multiple sensors in a grid covering Earth's surface across 37 pressure levels (1 to $\SI{1000}{\hecto\pascal}$). Measurements are sampled every six hours daily. For the analysis, we use ECMWF tools with inputs like the HAWC observatory’s coordinates (N 18° 59’ 39’’, W 97° 18’ 24’’, 4089 m elevation) and points along the horizontal trajectory. The tool returns the nearest grid point (e.g., Lat: 18.75, Long: 262.5 for HAWC, 33.9 km away) and four surrounding points, typically spaced ~40 km apart.

For the effective temperature calculation from the atmospheric data we follow the procedure in Ref. \cite{minosfar}, adapted to a horizontal trajectory for our case. The effective temperature, expessed as a weighted average, is:
\begin{align}
T_\mathrm{eff} = \frac{\int_{0}^{\infty} \dd{X} T(X)W(X)}
{\int_{0}^{\infty} \dd{X} W(X)}, 
\end{align}
where the weigths
\begin{align}
W^{\pi,K}(X) =
 \frac{  (1-X/ \Lambda_{\pi,K} ')^2  e^{-X/\Lambda_{\pi,K}} A^1_{\pi,K}
}{\gamma +\left( \gamma + 1 \right) B^1_{\pi,K} K(X)\left(E_\mathrm{th}\cos\theta / 
\epsilon_{\pi,K}\right)^2},
\end{align}
are dependent on parameters such as the meson masses, the attenuation lenghts $\Lambda$, the muon spectral index $\gamma$, the critical energies $\epsilon$ and the parameters $A$ and $B$, which describe the meson production in the forward region and the relative atmospheric attenuation, respectively.

\section{Simulations}
The MCEq software \cite{fedynitch2015} solves cascade equations numerically, offering efficiency compared to traditional Monte Carlo methods. We used Sibyll 2.3d, QGSJet-II-04, and EPOS-LHC hadronic models with HillasGaisser2012 H3a primary cosmic ray flux. Custom atmospheric profiles were implemented from the ECMWF data.

\subsection{Atmospheric profiles generation}
We developed an atmospheric profile generator based on CORSIKA's \texttt{gdastool} \cite{mitra2020}, producing profiles compatible with air-shower simulation softwares. The tool considers the 37 ECMWF pressure levels. Fixing values for pressure $P(h)$, temperature $T(h)$, and relative humidity $\text{relh}(h)$, the atmospheric density $\rho(h)$ is derived via the ideal gas law. Atmospheric depth $X(h)$ is subsequently computed using a multi stage fit. We make use of the 5-layer model of the atmosphere, where depth follows exponential functions in the lower four layers:
\begin{align}
X(h) = a_i +b_i e^{-h/c_i},\quad i=1,2,3,4
\end{align}
with linear approximation above \SI{100}{\kilo\meter}:
\begin{align}
X(h) = a_5 - \frac{b_5}{c_5}h.
\end{align}
The parameters $a_i$, $b_i$ and $c_i$ are optimized through nonlinear fitting with layer continuity constraints.

As a first approximation, we generated isothermal profiles ($\pm10\%$ from mean) to be able to isolate temperature effects. Since the majority of muons are produced in the upper atmosphere, where temperature gradients are relatively small compared to lower altitudes, an isothermal treatment captures the dominant thermal effects on pion and kaon decay probabilities. Then, we implemented the realistic ECMWF profiles at HAWC location, covering a range of effective temperature variations. This allows us to directly validate temperature effects against experimental muon horizontal flux measurements.

\subsection{Simulation results}
\begin{figure}[h]
\begin{subfigure}[t]{0.49\textwidth}
\centering
\includegraphics[width=0.9\textwidth]{ 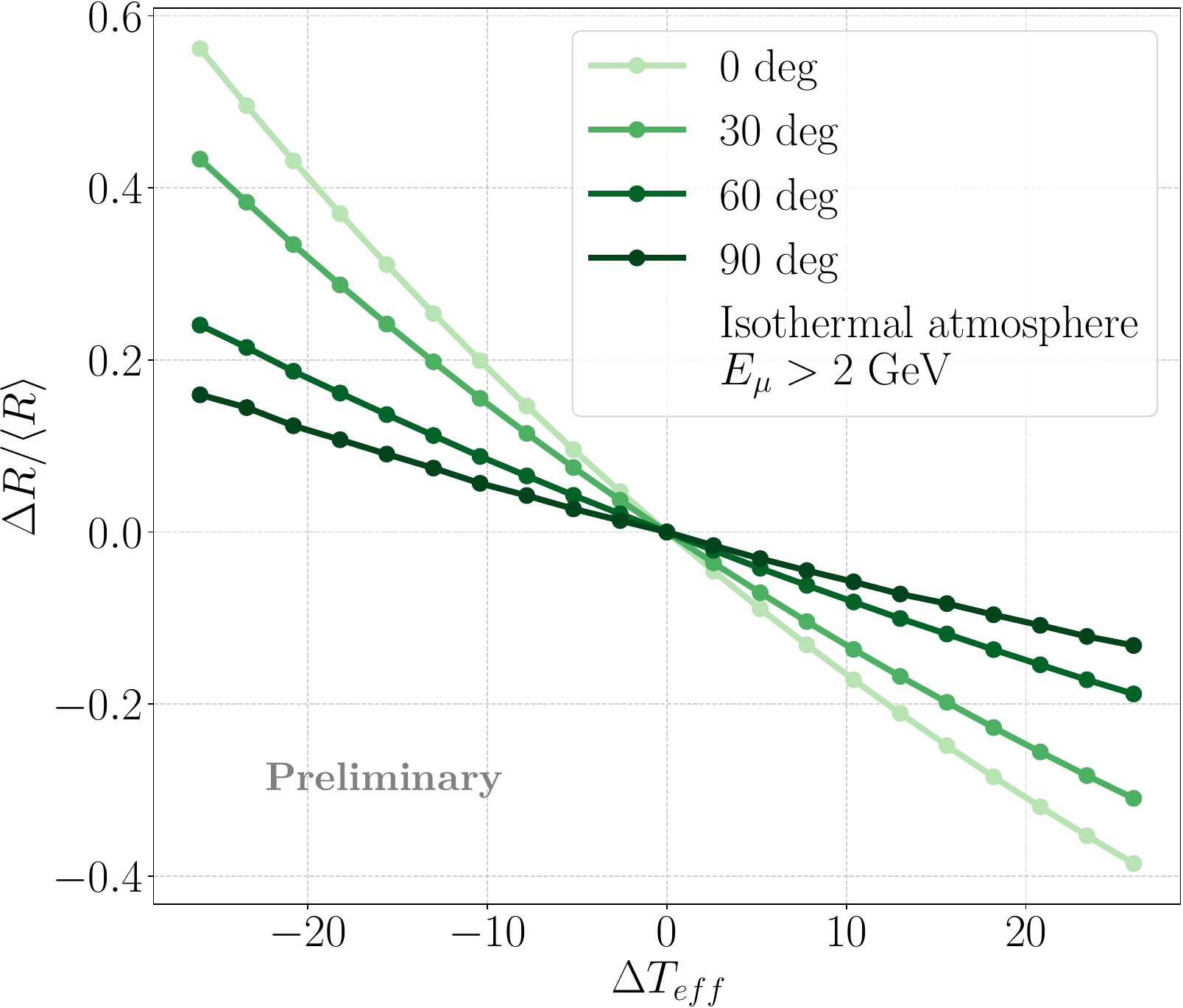}
\caption{Simulated muon rate vs. atmospheric temperature (in the isothermal approximation) for different zenith angles.}
\label{fig:corr-zenith}
\end{subfigure}
\hfill
\begin{subfigure}[t]{0.49\textwidth}
\centering
\includegraphics[width=0.9\textwidth]{ 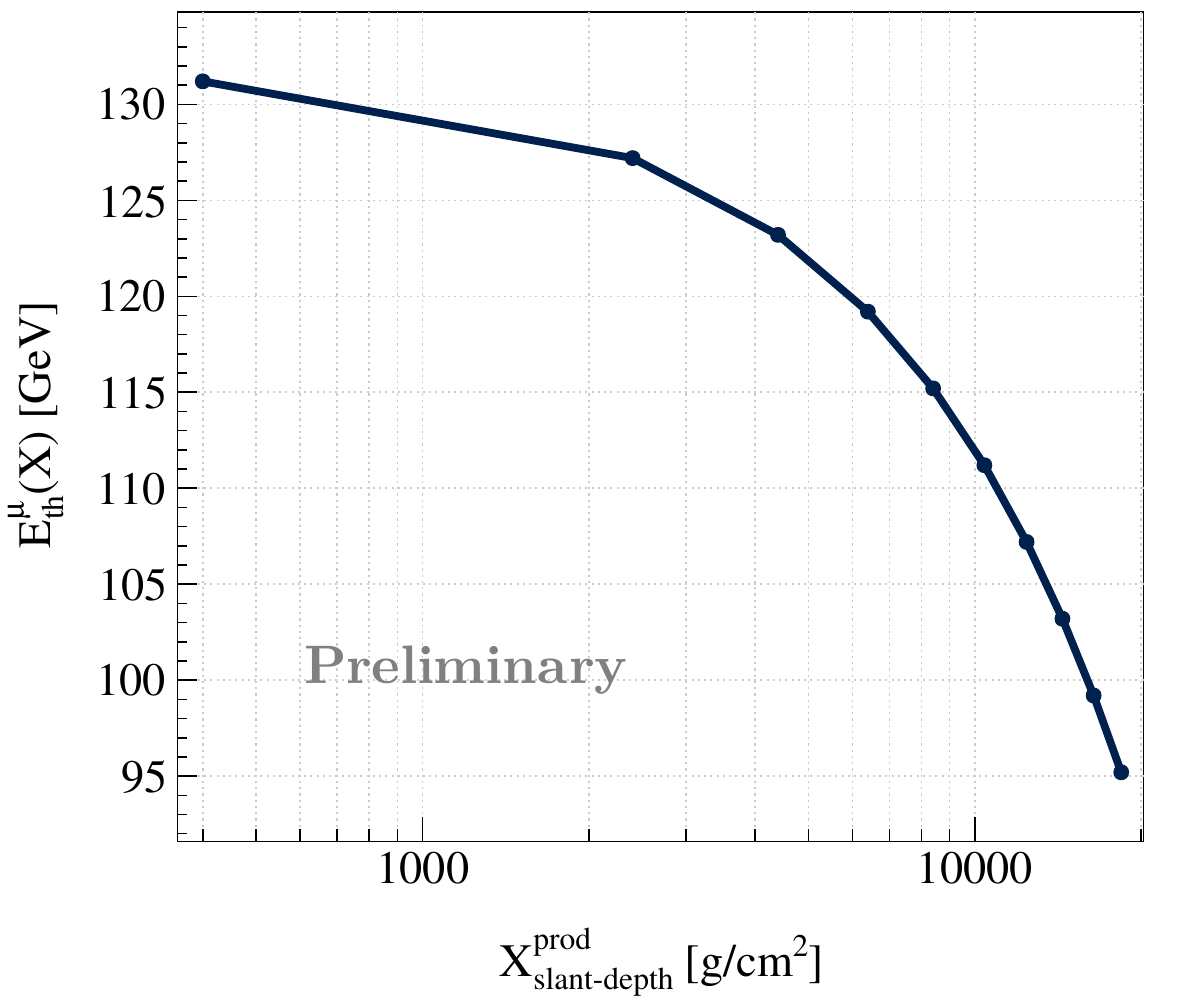}
\caption{Minimum muon energy required at production depth $X$ to reach HAWC with $E_\mu \geq 2$ GeV.}
\label{fig:muon-Eth}
\end{subfigure}
\caption{}
\end{figure}

With the isothermal appoximation of the atmosphere and a muon energy threshold of 2 GeV, we performed the simulations varying the primary zenith angle, shown in Fig.~\ref{fig:corr-zenith}. In general, we observe an anticorrelation between the muon rate and the temperature. However, this trend weaken progressively with increasing zenith angle, approaching a slope near zero for completely horizontal flux, just as we expected from the experimental results.


To properly estimate the energy threshold to be used in the integration for calculating the muon rate, it is important to account for the energy losses that muons experience as they propagate through the atmosphere. We assume a constant average energy loss rate of $\SI{2}{\MeV}/\si{\gram\per\centi\meter\squared}$ for muons traversing the atmosphere \cite{gaisser2016}. Using this approximation, we can estimate the energy a muon must have at the point of production (at slant depth $X$) in order to reach the observation level at depth $X_{obs}$ with sufficient energy to pass the detector threshold. The effective threshold energy at production depth is then given by
\begin{align}
E_{th}(X) = E_{th,obs} + \SI{2}{\MeV}/\si{\gram\per\centi\meter\squared} \qty(\frac{X_{obs}}{\cos\theta} - X),
\end{align}
where $E_{th}(X)$ is the required initial energy at slant depth $X$, $\theta$ is the zenith angle of the muon’s trajectory, and $E_{th,obs} \sim \SI{2}{\GeV}$ is the energy threshold at HAWC’s altitude. Figure~\ref{fig:muon-Eth} shows the resulting threshold energy $E_{th}(X)$ as a function of the slant depth, calculated using this expression. 

\begin{figure}[h]
\centering
\includegraphics[width=0.6\textwidth]{ 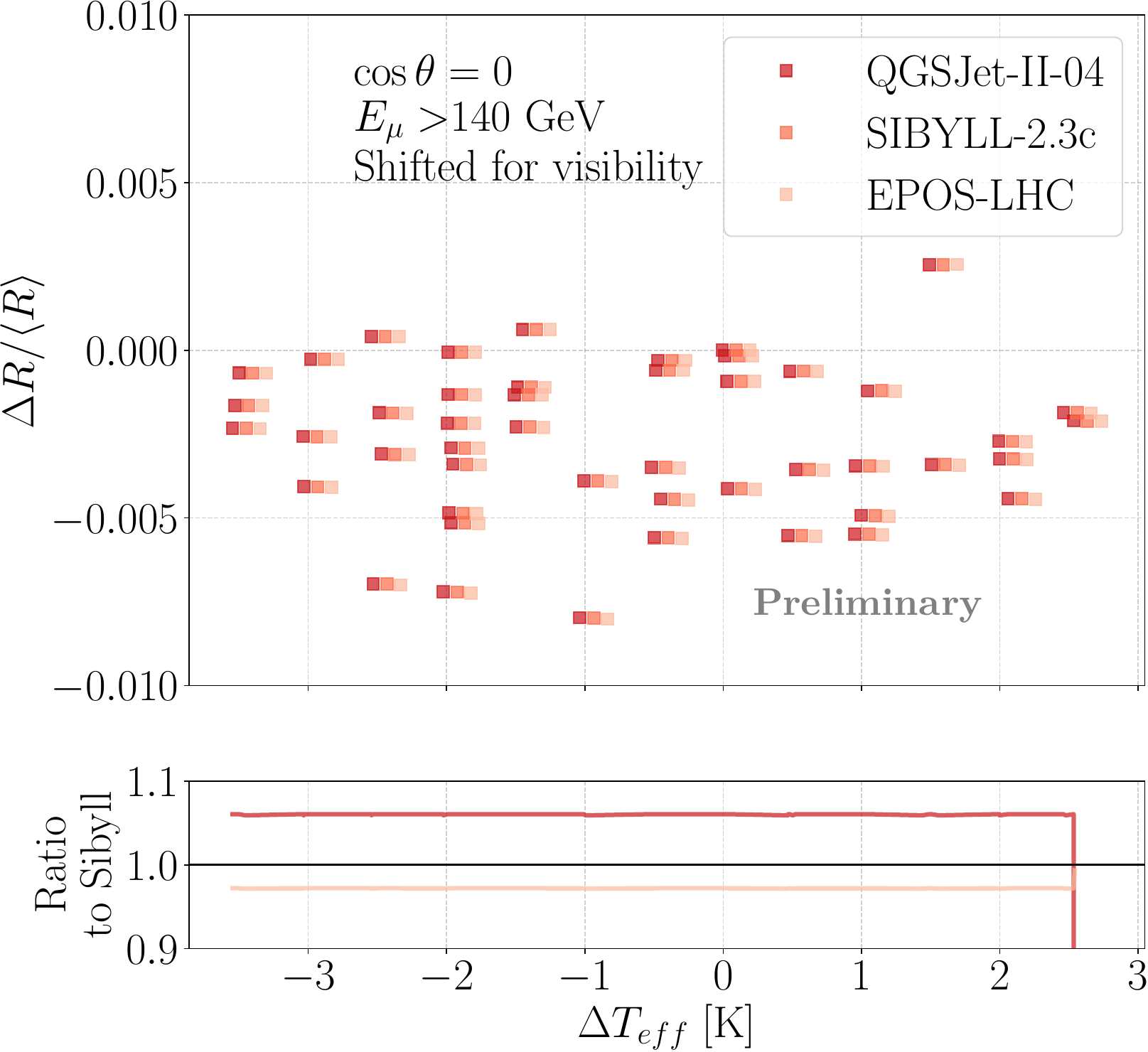}
\caption{Comparison of hadronic interaction models (Sibyll 2.3, QGSJet-II-04, EPOS-LHC) for horizontal muon rate vs. effective temperature. Points are horizontally offset for clarity.}
\label{fig:corr-hadmodels}
\end{figure}

To evaluate the dependence of our results on the choice of hadronic interaction model, we performed the simulations using three different models. For these, we use the highest muon energy threshold considered ($\SI{140}{\GeV}$), as well as the actual atmospheric profiles from ECMWF data, corresponding to different effective temperatures. The results, shown in Figure~\ref{fig:corr-hadmodels}, reveal no significant differences between the models in the resulting rate variations at observation level. For visibility, the points are slightly shifted along horizontal axis by $0.1$ and $\SI{0.2}{\kelvin}$ for Sibyll 2.3c and EPOS-LHC, respectively, since the models produce virtually identical results. This reinforces our conclusion that the observed weak correlation for horizontal muons is a general feature of the production and transport mechanisms in the temperature variation range, and not something caused by the choice of hadronic model.

\vspace{-5mm}
\subsubsection*{Comparison with data}
\begin{figure}[h!]
\centering
\includegraphics[width=0.6\textwidth]{ 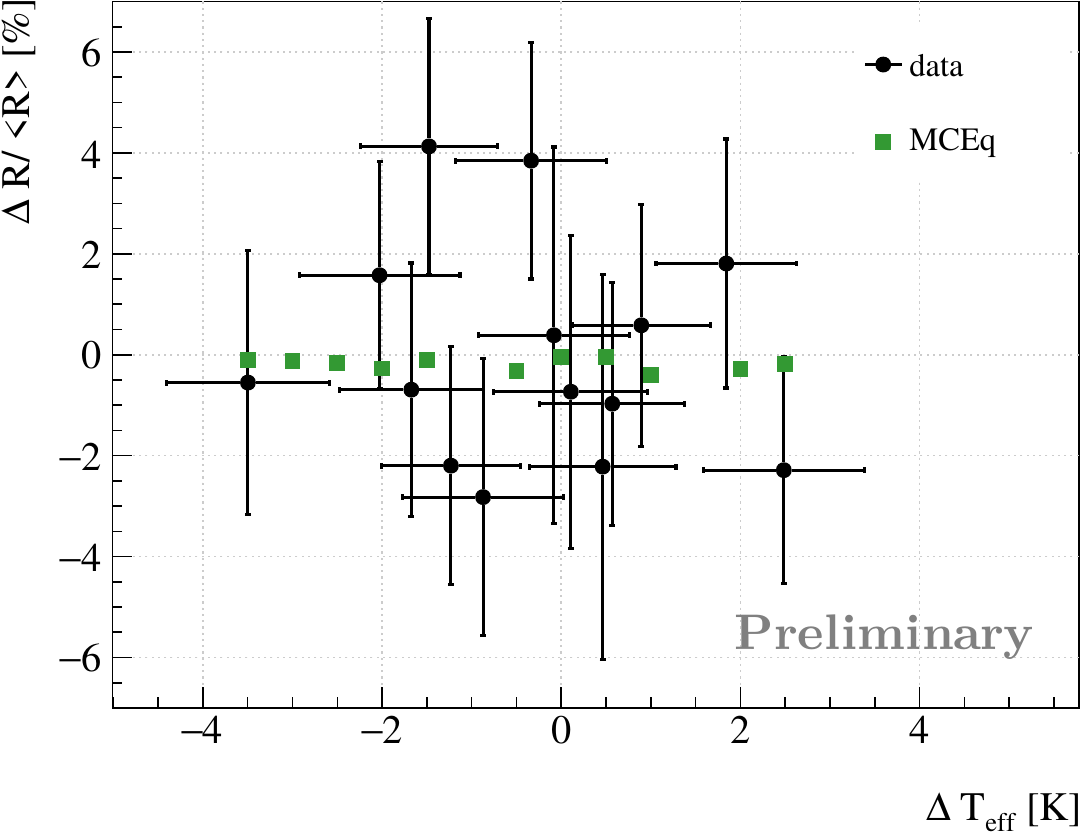}
\caption{Comparison between HAWC data (black points) and MCEq simulations with Sibyll 2.3 (green squares) for horizontal muon rate vs. effective temperature. Both data sets show the absence of significant correlation.}
\label{fig:mceq-obs}
\end{figure}

Finally, we present a direct comparison between the simulation results and experimental data in Figure~\ref{fig:mceq-obs}. Despite the large uncertainties present in the data, we find the agreement between the MCEq predictions and the observed horizontal muon rates to be reasonable. Importantly, the MCEq results show no significant correlation between the horizontal muon rate and the effective atmospheric temperature, consistent with the behavior observed in the HAWC measurements. This consistency further supports our conclusion that temperature induced variations in the horizontal muon flux are negligible, both in simulations and in actual data.\vspace{-0.5mm}

\newpage
\section{Summary}
\vspace{-2mm}
We have presented an analysis of the horizontal muon flux detected by the HAWC observatory, using an extended data sample based on the Earth-skimming neutrino search~\cite{hawc-neutrinos}. To analyze the observed muon rates, we made use of realistic atmospheric profiles derived from ECMWF data and calculated the corresponding effective temperatures following the method described in~\cite{minosfar}. Numerical simulations were performed using the MCEq framework~\cite{fedynitch2015} with various hadronic interaction models considering both isothermal and realistic atmospheric conditions. The results, from both simulation and experimental data, suggest that variations in the horizontal muon flux at HAWC due to atmospheric temperature variations are negligible. This work represents the first reported search for seasonal variations in the horizontal muon rate at high altitude.

\section*{Acknowledgments}
\vspace{-2mm}
\scriptsize
This work was supported by CONACYT-SECIHTI, Mexico, grant CF-2023-I-645 and the postgraduate scholarship program, and by PAPIIT-UNAM grant IN102223. 

We acknowledge the support from: the US National Science Foundation (NSF); the US Department of Energy Office of High-Energy Physics; the Laboratory Directed Research and Development (LDRD) program of Los Alamos National Laboratory; Consejo Nacional de Ciencia y Tecnolog\'{i}a (CONACyT), M\'{e}xico, grants LNC-2023-117, 271051, 232656, 260378, 179588, 254964, 258865, 243290, 132197, A1-S-46288, A1-S-22784, CF-2023-I-645, CBF2023-2024-1630, c\'{a}tedras 873, 1563, 341, 323, Red HAWC, M\'{e}xico; DGAPA-UNAM grants IG101323, IN111716-3, IN111419, IA102019, IN106521, IN114924, IN110521 , IN102223; VIEP-BUAP; PIFI 2012, 2013, PROFOCIE 2014, 2015; the University of Wisconsin Alumni Research Foundation; the Institute of Geophysics, Planetary Physics, and Signatures at Los Alamos National Laboratory; Polish Science Centre grant, 2024/53/B/ST9/02671; Coordinaci\'{o}n de la Investigaci\'{o}n Cient\'{i}fica de la Universidad Michoacana; Royal Society - Newton Advanced Fellowship 180385; Gobierno de España and European Union-NextGenerationEU, grant CNS2023- 144099; The Program Management Unit for Human Resources \& Institutional Development, Research and Innovation, NXPO (grant number B16F630069); Coordinaci\'{o}n General Acad\'{e}mica e Innovaci\'{o}n (CGAI-UdeG), PRODEP-SEP UDG-CA-499; Institute of Cosmic Ray Research (ICRR), University of Tokyo. H.F. acknowledges support by NASA under award number 80GSFC21M0002. C.R. acknowledges support from National Research Foundation of Korea (RS-2023-00280210). We also acknowledge the significant contributions over many years of Stefan Westerhoff, Gaurang Yodh and Arnulfo Zepeda Dom\'inguez, all deceased members of the HAWC collaboration. Thanks to Scott Delay, Luciano D\'{i}az and Eduardo Murrieta for technical support.

%
\bibliographystyle{JHEP} 
\bibliography{horizontalmuons} 

\section*{Full Author List: \ HAWC Collaboration}
\vspace{-1cm}
\noindent

\vskip1cm
\noindent

R. Alfaro$^{1}$,
C. Alvarez$^{2}$,
A. Andrés$^{3}$,
E. Anita-Rangel$^{3}$,
M. Araya$^{4}$,
J.C. Arteaga-Velázquez$^{5}$,
D. Avila Rojas$^{3}$,
H.A. Ayala Solares$^{6}$,
R. Babu$^{7}$,
P. Bangale$^{8}$,
E. Belmont-Moreno$^{1}$,
A. Bernal$^{3}$,
K.S. Caballero-Mora$^{2}$,
T. Capistrán$^{9}$,
A. Carramiñana$^{10}$,
F. Carreón$^{3}$,
S. Casanova$^{11}$,
S. Coutiño de León$^{12}$,
E. De la Fuente$^{13}$,
D. Depaoli$^{14}$,
P. Desiati$^{12}$,
N. Di Lalla$^{15}$,
R. Diaz Hernandez$^{10}$,
B.L. Dingus$^{16}$,
M.A. DuVernois$^{12}$,
J.C. Díaz-Vélez$^{12}$,
K. Engel$^{17}$,
T. Ergin$^{7}$,
C. Espinoza$^{1}$,
K. Fang$^{12}$,
N. Fraija$^{3}$,
S. Fraija$^{3}$,
J.A. García-González$^{18}$,
F. Garfias$^{3}$,
N. Ghosh$^{19}$,
A. Gonzalez Muñoz$^{1}$,
M.M. González$^{3}$,
J.A. Goodman$^{17}$,
S. Groetsch$^{19}$,
J. Gyeong$^{20}$,
J.P. Harding$^{16}$,
S. Hernández-Cadena$^{21}$,
I. Herzog$^{7}$,
D. Huang$^{17}$,
P. Hüntemeyer$^{19}$,
A. Iriarte$^{3}$,
S. Kaufmann$^{22}$,
D. Kieda$^{23}$,
K. Leavitt$^{19}$,
H. León Vargas$^{1}$,
J.T. Linnemann$^{7}$,
A.L. Longinotti$^{3}$,
G. Luis-Raya$^{22}$,
K. Malone$^{16}$,
O. Martinez$^{24}$,
J. Martínez-Castro$^{25}$,
H. Martínez-Huerta$^{30}$,
J.A. Matthews$^{26}$,
P. Miranda-Romagnoli$^{27}$,
P.E. Mirón-Enriquez$^{3}$,
J.A. Montes$^{3}$,
J.A. Morales-Soto$^{5}$,
M. Mostafá$^{8}$,
M. Najafi$^{19}$,
L. Nellen$^{28}$,
M.U. Nisa$^{7}$,
N. Omodei$^{15}$,
E. Ponce$^{24}$,
Y. Pérez Araujo$^{1}$,
E.G. Pérez-Pérez$^{22}$,
Q. Remy$^{14}$,
C.D. Rho$^{20}$,
D. Rosa-González$^{10}$,
M. Roth$^{16}$,
H. Salazar$^{24}$,
D. Salazar-Gallegos$^{7}$,
A. Sandoval$^{1}$,
M. Schneider$^{1}$,
G. Schwefer$^{14}$,
J. Serna-Franco$^{1}$,
A.J. Smith$^{17}$
Y. Son$^{29}$,
R.W. Springer$^{23}$,
O. Tibolla$^{22}$,
K. Tollefson$^{7}$,
I. Torres$^{10}$,
R. Torres-Escobedo$^{21}$,
R. Turner$^{19}$,
E. Varela$^{24}$,
L. Villaseñor$^{24}$,
X. Wang$^{19}$,
Z. Wang$^{17}$,
I.J. Watson$^{29}$,
H. Wu$^{12}$,
S. Yu$^{6}$,
S. Yun-Cárcamo$^{17}$,
H. Zhou$^{21}$,

\vskip1cm
\noindent

$^{1}$Instituto de F\'{i}sica, Universidad Nacional Autónoma de México, Ciudad de Mexico, Mexico,
$^{2}$Universidad Autónoma de Chiapas, Tuxtla Gutiérrez, Chiapas, México,
$^{3}$Instituto de Astronom\'{i}a, Universidad Nacional Autónoma de México, Ciudad de Mexico, Mexico,
$^{4}$Universidad de Costa Rica, San José 2060, Costa Rica,
$^{5}$Universidad Michoacana de San Nicolás de Hidalgo, Morelia, Mexico,
$^{6}$Department of Physics, Pennsylvania State University, University Park, PA, USA,
$^{7}$Department of Physics and Astronomy, Michigan State University, East Lansing, MI, USA,
$^{8}$Temple University, Department of Physics, 1925 N. 12th Street, Philadelphia, PA 19122, USA,
$^{9}$Universita degli Studi di Torino, I-10125 Torino, Italy,
$^{10}$Instituto Nacional de Astrof\'{i}sica, Óptica y Electrónica, Puebla, Mexico,
$^{11}$Institute of Nuclear Physics Polish Academy of Sciences, PL-31342 11, Krakow, Poland,
$^{12}$Dept. of Physics and Wisconsin IceCube Particle Astrophysics Center, University of Wisconsin{\textemdash}Madison, Madison, WI, USA,
$^{13}$Departamento de F\'{i}sica, Centro Universitario de Ciencias Exactase Ingenierias, Universidad de Guadalajara, Guadalajara, Mexico, 
$^{14}$Max-Planck Institute for Nuclear Physics, 69117 Heidelberg, Germany,
$^{15}$Department of Physics, Stanford University: Stanford, CA 94305–4060, USA,
$^{16}$Los Alamos National Laboratory, Los Alamos, NM, USA,
$^{17}$Department of Physics, University of Maryland, College Park, MD, USA,
$^{18}$Tecnologico de Monterrey, Escuela de Ingenier\'{i}a y Ciencias, Ave. Eugenio Garza Sada 2501, Monterrey, N.L., Mexico, 64849,
$^{19}$Department of Physics, Michigan Technological University, Houghton, MI, USA,
$^{20}$Department of Physics, Sungkyunkwan University, Suwon 16419, South Korea,
$^{21}$Tsung-Dao Lee Institute \& School of Physics and Astronomy, Shanghai Jiao Tong University, 800 Dongchuan Rd, Shanghai, SH 200240, China,
$^{22}$Universidad Politecnica de Pachuca, Pachuca, Hgo, Mexico,
$^{23}$Department of Physics and Astronomy, University of Utah, Salt Lake City, UT, USA, 
$^{24}$Facultad de Ciencias F\'{i}sico Matemáticas, Benemérita Universidad Autónoma de Puebla, Puebla, Mexico, 
$^{25}$Centro de Investigaci\'on en Computaci\'on, Instituto Polit\'ecnico Nacional, M\'exico City, M\'exico,
$^{26}$Dept of Physics and Astronomy, University of New Mexico, Albuquerque, NM, USA,
$^{27}$Universidad Autónoma del Estado de Hidalgo, Pachuca, Mexico,
$^{28}$Instituto de Ciencias Nucleares, Universidad Nacional Autónoma de Mexico, Ciudad de Mexico, Mexico, 
$^{29}$University of Seoul, Seoul, Rep. of Korea,
$^{30}$Departamento de Física y Matemáticas, Universidad de Monterrey, Av.~Morones Prieto 4500, 66238, San Pedro Garza Garc\'ia NL, M\'exico
\end{document}